\pdfoutput=1

\documentclass[11pt]{article}

\usepackage[]{acl}

\usepackage{times}
\usepackage{latexsym}

\usepackage[T1]{fontenc}

\usepackage[utf8]{inputenc}

\usepackage{microtype}

\usepackage{inconsolata}
\usepackage{graphicx}
\usepackage{amsfonts}
\usepackage{amsmath}
\usepackage{adjustbox}
\usepackage{multirow}

\title{MobileSpeech: A Fast and High-Fidelity Framework \\ for Mobile Zero-Shot Text-to-Speech}

\author{Shengpeng Ji\thanks{Equal contribution.}$^{1,2}$, Ziyue Jiang\footnotemark[1]$^{1}$, Hanting Wang$^{1}$, Jialong Zuo$^{1}$, Zhou Zhao\thanks{Corresponding author.}$^{1,2}$  \\
  Zhejiang University$^{1}$ \\
  Shanghai AI Laboratory$^{2}$\\
  \texttt{shengpengji,zhaozhou@zju.edu.cn}  }

\begin{document}
\maketitle
\begin{abstract}
Zero-shot text-to-speech (TTS) has gained significant attention due to its powerful voice cloning capabilities, requiring only a few seconds of unseen speaker voice prompts. However, all previous work has been developed for cloud-based systems. Taking autoregressive models as an example, although these approaches achieve high-fidelity voice cloning, they fall short in terms of inference speed, model size, and robustness. Therefore, we propose MobileSpeech, which is a fast, lightweight, and robust zero-shot text-to-speech system based on mobile devices for the first time. Specifically: 1) leveraging discrete codec, we design a parallel speech mask decoder module called SMD, which incorporates hierarchical information from the speech codec and weight mechanisms across different codec layers during the generation process. Moreover, to bridge the gap between text and speech, we introduce a high-level probabilistic mask that simulates the progression of information flow from less to more during speech generation. 2) For speaker prompts, we extract fine-grained prompt duration from the prompt speech and incorporate text, prompt speech by cross attention in SMD. We demonstrate the effectiveness of MobileSpeech on multilingual datasets at different levels, achieving state-of-the-art results in terms of generating speed and speech quality. MobileSpeech achieves RTF of 0.09 on a single A100 GPU and we have successfully deployed MobileSpeech on mobile devices. Audio samples are available at \url{https://mobilespeech.github.io/} .
\end{abstract}

\section{Introduction}

In recent years, remarkable progress has been made in the development of text-to-speech (TTS) technology \cite{ren2019fastspeech,vits,fastdiff,styletts2}. With advancements in this field, current state-of-the-art TTS systems have demonstrated the ability to generate high-quality voices of unseen speakers while maintaining consistency in tone and rhythm \cite{adaspeech,yourtts}. However, these systems are often trained on datasets of only a few hundred hours and often require fine-tuning with dozens of sentences during the inference stage, which significantly limits the generative capabilities of the models and is time-consuming and not user-friendly.

With the recent development of large-scale language models \cite{gpt,llama} and the increase in training data for speech synthesis from hundreds of hours \cite{libritts} to tens of thousands of hours \cite{librilight} , current TTS models exhibit powerful zero-shot capabilities, enabling the cloning of unseen speakers' voices in just a few seconds \cite{valle,vallex,naturalspeech2,le2023voicebox,megatts,speartts,soundstorm,make-a-voive,uniaudio} . However, these models primarily focus on in-context abilities and fail to address issues related to inference speed and model deployment parameters. Additionally, all existing zero-shot TTS models are cloud-based and lack a mobile-based deployment approach. MobileSpeech is the first zero-shot TTS synthesis system that can be deployed on mobile devices. Firstly we think deployable mobile zero-shot TTS models aim to achieve the following goals:

\begin{itemize}
    \item \textbf{Fast:} through our experiments, we have observed that the real-time factor (RTF) on a single mobile device often exceeds 8-10 times that of an A100 RTF for the same text. Therefore, the inference speed of the zero-shot TTS model must be significantly faster.
    \item \textbf{Lightweight:} to enable deployment on mobile or edge devices, the model size should be small, and the runtime memory footprint should be minimal.
    \item \textbf{High similarity and diversity:} zero-shot TTS system should be able to clone timbre and prosody with just a few seconds of prompts, yielding diverse speech even when provided with identical text.
    \item \textbf{High quality:} to enhance the naturalness of synthesized speech, the model should pay attention to details such as frequency bins between adjacent harmonics and possess powerful duration modeling capabilities.
    \item \textbf{Robustness:} a highly robust TTS system is essential. A zero-shot TTS system should minimize occurrences of missing or repeated words.
\end{itemize}

To achieve these objectives, we have made our best efforts to reproduce several state-of-the-art generative models, which will be elaborated in the subsequent section on related work. We found  current work is not applicable to mobile devices. Ultimately, based on the Natspeech framework \cite{ren2019fastspeech,fastspeech2} and the Mask generation module \cite{maskgit,soundstorm}, we additionally 
design the SMD module and the speaker prompt component, taking into account the discrete codec architecture of speech. In summary, MobileSpeech contributes as follows:

\begin{itemize}
    \item MobileSpeech is the first zero-shot TTS synthesis system that can be deployed on mobile devices. MobileSpeech achieves a good balance according to the above five evaluation metrics.
    \item In MobileSpeech, we additionally design the SMD module based on the hierarchical token structure of discrete speech codecs and design the Speaker Prompt module to maintain high similarity, high quality, low latency.
    \item Training MobileSpeech-english for 580 hours achieved SOTA inference speed and comparable audio quality. Training MobileSpeech-chinese for 40,000 hours resulted in SOTA performance in all aspects.
    \item MobileSpeech has been successfully deployed on mobile phones and is expected to be used by several hundred thousand users. 
\end{itemize}

\section{Related Work}

\subsection{Zero-shot TTS}
Zero-shot speech synthesis refers to the ability to synthesize the voice of an unseen speaker based solely on a few seconds of audio prompt, also known as voice cloning. Due to its impressive performance, it has gained significant attention recently.
Previous approaches in the field of voice cloning can be classified into speaker adaptation \cite{styletokens,adaspeech,metatts} and speaker encoding methods \cite{yourtts,anyspeaker2,anyspeaker}. These models are often trained on smaller datasets and require a substantial amount of data for fine-tuning or employ highly complex encoders. These limitations greatly restrict the audio quality and further usage. 

\begin{figure*}[t]
\centering
\includegraphics[height=7.5cm, width=16cm]{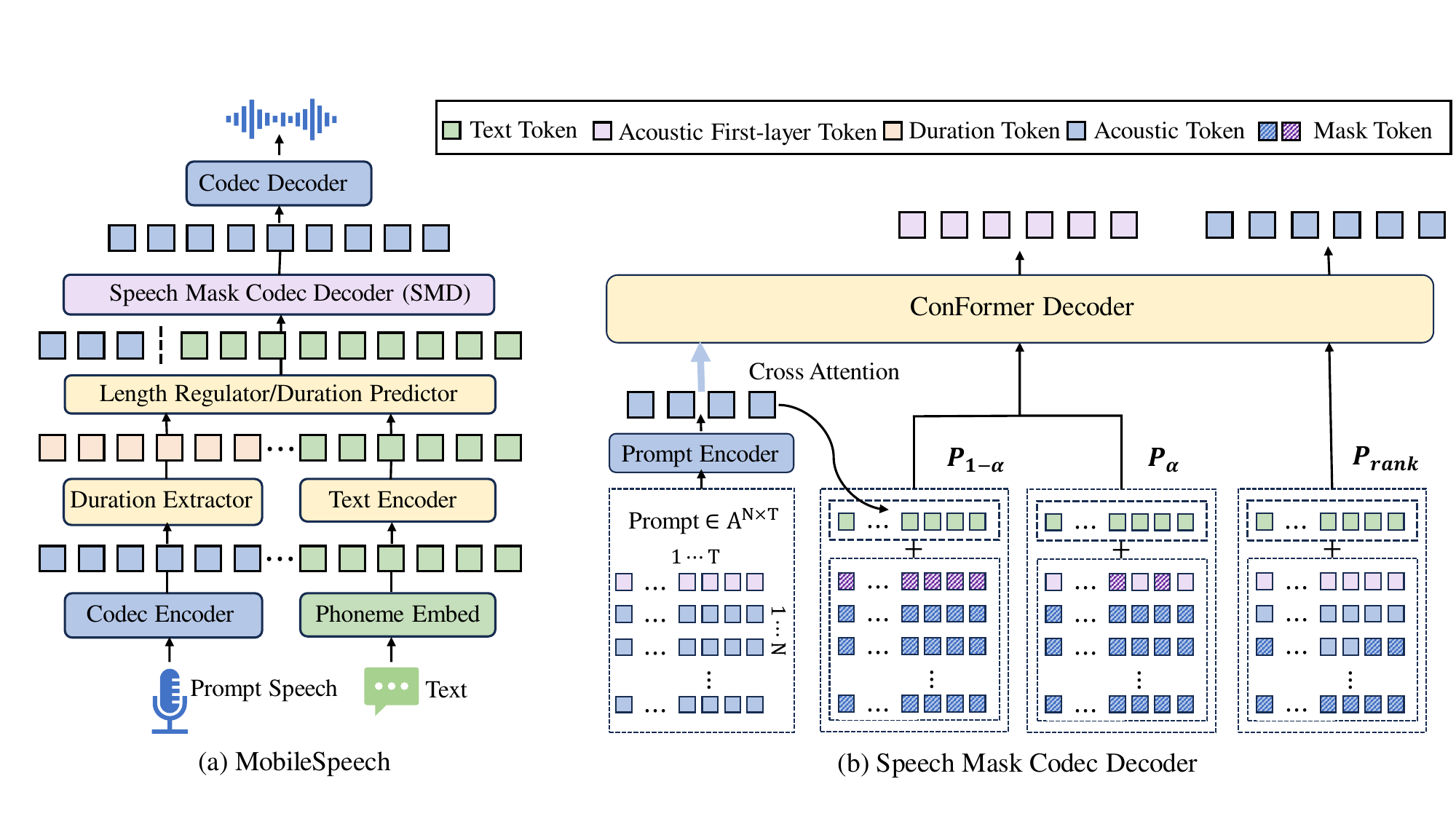}

\caption{The overall architecture for MobileSpeech. In Figure (a) the Duration Extractor module is responsible for extracting prompt duration tokens from prompt acoustic tokens. SMD represents the generative module for target acoustic tokens. In Figure (b), we provide a detailed depiction of the multi-channel training process employed by the SMD module.}
\label{figurejsp1}
\end{figure*}

In recent months, with the advancement of generative large-scale models, a plethora of outstanding works have emerged. VALL-E \cite{valle} leverages discrete codec representations and combines autoregressive and non-autoregressive models in a cascaded manner, preserving the powerful contextual capabilities of language models. It can clone the voice of a target speaker with just a 3-second audio prompt. VALL-E X \cite{vallex} extends zero-shot TTS to multiple languages based on the cascaded structure of VALL-E. NaturalSpeech 2 \cite{naturalspeech2} employs continuous vectors instead of discrete neural codec tokens and introduces in-context learning to a latent diffusion model. SpearTTS \cite{speartts} and Make-a-Voice \cite{make-a-voive} utilize semantic tokens to reduce the gap between text and acoustic features. VoiceBox \cite{le2023voicebox} is a non-autoregressive flow-matching model trained to infill speech, given audio context and text. Megatts \cite{megatts,megatts2} , on the other hand, utilizes traditional mel-spectrograms, decoupling timbre and prosody and further modeling the prosody using an autoregressive approach.

However, none of the aforementioned works take into account the model's speed and lightweight nature. It is well-known that autoregressive models often require large memory footprints and have slower iteration speeds, and diffusion models often require hundreds of sampling steps. Some work introduce a semantic token cascade structure, which will introduce more time overhead. However, compared to codecs, certain approaches based on mel-spectrogram exhibit lower generation diversity and audio quality \cite{vocos}. MobileSpeech, on the other hand, is the first zero-shot TTS system that can be deployed on mobile devices, addressing the fast speed,lightweight, high quality requirements.

\subsection{Generative models}
Generative models have been extensively utilized in various domains by employing techniques such as language models \cite{audiolm,audiogen,textrolspeech,valle,vallex}, Variational Autoencoders (VAE) \cite{portaspeech,vae1}, Generative Adversarial Networks (GAN) \cite{gan1,gan2}, Normalizing Flow \cite{flow1,flow2}, Mask Autoencoders \cite{maskgit}and diffusion models \cite{fastdiff}. Considering the challenge of balancing perceptual quality and inference speed in zero-shot TTS, as well as the recent success of parallel decoding in text \cite{textmask}, image \cite{maskgit}, audio \cite{soundstorm}, and video \cite{videomask} generation tasks, MobileSpeech has adopted masked parallel generation as the audio synthesis approach. Furthermore, MobileSpeech has additionally designed the Speech Codec Mask Decoder (SMD) module and Speaker prompt based on the unique characteristics of acoustic tokens and the specific requirements of the zero-shot TTS task, aiming to address the aforementioned trade-off. In Appendix \ref{appendixA}, we present the detail of discrete acoustic codec and in Appendix \ref{appendixB} we present the distinctions between mask-based generative models and other generative models in modeling discrete codecs.

\section{MobileSpeech}
In this section, we will first introduce the overall structure of MobileSpeech, followed by a detailed focus on the Speech Mask Codec Decoder module and the Speaker Prompt module. Furthermore, we will proceed to elaborate on the specific intricacies of MobileSpeech's training and inference processes, with the explicit details of the model being presented in Appendix \ref{appendixb}.
\subsection{Overall}
As shown in Figure \ref{figurejsp1}, MobileSpeech employs FastSpeech 2 as our baseline. The input text is processed to extract phoneme sequence, which are then passed through a text encoder to obtain the corresponding target text representation. Simultaneously, the input consists of speech prompts based on a discrete codec \cite{encodec}. The prompt encoder compresses the channel dimension of the prompt codec acoustic tokens to one dimension. We specifically extract fine-grained prompt duration tokens from the prompt codec acoustic tokens and combine them with the target text, which are then fed into the duration prediction module and length regularizer to obtain an expanded target text. The Speech Mask Codec Decoder serves as the core module. During the training process, we randomly select a portion of the aligned target text as the prompt to guide the generation of the remaining text. Based on the masking and parallel generation using RVQ features of the speech codec, we design a multi-channel training approach, priority training, and a mechanism that relies solely on text to generate acoustic tokens. Finally, the acoustic tokens are transformed into the final speech output using a pre-trained codec Tokenizer \cite{vocos}.

\subsection{Speech Codec Mask Decoder}
Based on paired text-speech data $\left \{ X,Y \right \} $, where $X=\left \{ x_{1},x_{2},x_{3},\cdots, x_{T}\right \} $ represents the aligned text and $Y$ denotes the representation of speech through a pre-trained discrete codec encoder \cite{encodec}, namely $Y=C_{1:T,1:N}\in  \mathbb{R}^{T\times N} $, we consider $T$ as the downsampled utterance length, which is equal to the length of the text. $N$ represents the number of channels for every frame. The row vector of each acoustic code matrix $C_{t,:}$ represents the eight codes for frame $t$, and the column vector of each acoustic code matrix $C_{:,j}$ represents the code sequence from the $j$-th codebook, where $j\in \left \{ 1,2,\cdots,N \right \} $ .

Notably, in contrast to models such as VALL-E \cite{valle}, it is acknowledged that obtaining prompt text in real-world scenarios is challenging, and to mitigate the impact of erroneous prompt text on the model. We have devised a text-agnostic speech prompt to guide the generation of target speech in MobileSpeech. Specifically, for a sequence with a time length of $T$, we randomly extract a segment of length $k$ at the beginning to simulate prompts of arbitrary lengths, where the value of $k$ falls within the range of:
\begin{equation}
    \frac{T}{3} \le k \le \frac{2\times T}{3} \label{equation1}
\end{equation}

By selecting the split point k, we obtain the corresponding prompt speech $Y_{prompt}$ and the target text $Y_{prompt}$:
\begin{equation}
    Y_{prompt} = C_{1:k,1:N} ,X_{target}=X_{k:T}
    \label{equaton2}
\end{equation}

Therefore, disregarding the codec channel count, the optimization objective of the Speech Codec Mask Decoder (SMD) module is to maximize the conditional probability function shown below:
\begin{equation}
    P(C_{k:T,1:N}\mid C_{1:k,1:N},X_{k:T};\theta)
\end{equation}

It is well known that the use of RVQ-based speech discretization tokens \cite{encodec,vocos} induces a hierarchical token structure, where tokens from finer RVQ levels contribute less to the perceptual quality. This allows for efficient factorization and approximation of the joint distribution of the token sequence. We believe that considering this structural aspect holds the most promise for future advancements in long-sequence audio modeling. Therefore, in MobileSpeech, as depicted in Figure \ref{figurejsp1} (b), we divide the generation process into two parts: the generation of the first channel and the generation of the remaining channels at the same time, denoted as:
\begin{equation}
    \begin{aligned}
        &P(C_{k:T,1:N}\mid C_{1:k,1:N},X_{k:T};\theta) \\  =&P(C_{k:T,1}\mid C_{1:k,1:N},X_{k:T};\theta)  \\        \times&\prod_{j=2}^{N}P(C_{k:T,j}\mid C_{1:k,1:N},X_{k:T};\theta) 
    \end{aligned}
\end{equation}

For the generation of the first channel $P(C_{k:T,1}\mid C_{1:k,1:N},X_{k:T};\theta)$, we employ a mask-based generative model as our parallel decoder. We sample the mask $M_{i}\in \left \{ 0,1 \right \} ^T$ according to a cosine schedule \cite{maskgit} for codec level $i$, specifically sampling the masking ratio $p=\cos(u)$ where $u\sim \mathcal{U}\left [  0,\frac{\pi }{2}  \right ] $. and the mask $M_{i} \sim  Bernoulli(p) $. Here, $M_{i}$ represents the portion to be masked in the $i$-th level, while $\bar{M_{i}} $ denotes the unmasked portion in the $i$-th level. The prediction of this portion is refined based on the prompt speech and the concat of target text and the unmasked portion of the first channel. Therefore, the prediction for this part can be specified as follows:
\begin{equation}
    P(M_{1}C_{k:T,1}\mid C_{1:k,1:N},X_{k:T},\bar{M_{1}}C_{k:T,1};\theta) 
\end{equation}

Due to the fact that in the inference process, we directly obtain the first-layer discrete tokens from the text, it often requires multiple iterations to iteratively select higher-confidence tokens to simulate the masked environment during training. In order to reduce the number of iterations, as shown in Figure \ref{figurejsp1} (b), MobileSpeech, during the training process, employs a certain probability to completely mask the acoustic tokens of the first channel. Additionally, completely masking the acoustic tokens of the first channel at different stages of model training can guide the model to generate from easier to harder instances, thus increasing the upper limit of the model. We control the two different ways of generating the first channel through a probability parameter $\alpha $, as shown below:

\begin{equation}
    \begin{aligned}
        &P(C_{k:T,1}\mid C_{1:k,1:N},X_{k:T};\theta) \\
        =\alpha P(M_{1}&C_{k:T,1}\mid C_{1:k,1:N},X_{k:T},\bar{M_{1}}C_{k:T,1};\theta)\\
        +(1-\alpha )&P(M_{1}C_{k:T,1}\mid C_{1:k,1:N},X_{k:T};\theta) 
    \end{aligned}
\end{equation}

For the generation of the remaining channels, similar to other models that randomly select a channel for training during the training process, we incorporate a weighting mechanism in the random selection process. We believe that the importance of channels decreases layer by layer. Therefore, we introduce the $P_{rank}$ mechanism in Appendix \ref{appendixc}. When predicting the acoustic token of the $j$-th channel, we utilize the unmasked portion of the $j$-th layer $\bar{M_{j}}$ and all acoustic tokens preceding the $j$-th layer for prediction, as shown below:
\begin{equation}
\resizebox{0.48\textwidth}{!}{$\displaystyle P(M_{j}C_{k:T,j}\mid C_{1:k,1:N},X_{k:T},C_{k:T,<j},\bar{M_{j}}C_{k:T,j};\theta)$}
\end{equation}


\subsection{Speaker Prompt}
\begin{figure}[htbp]
    \centering
    \includegraphics[height=4cm, width=8cm]{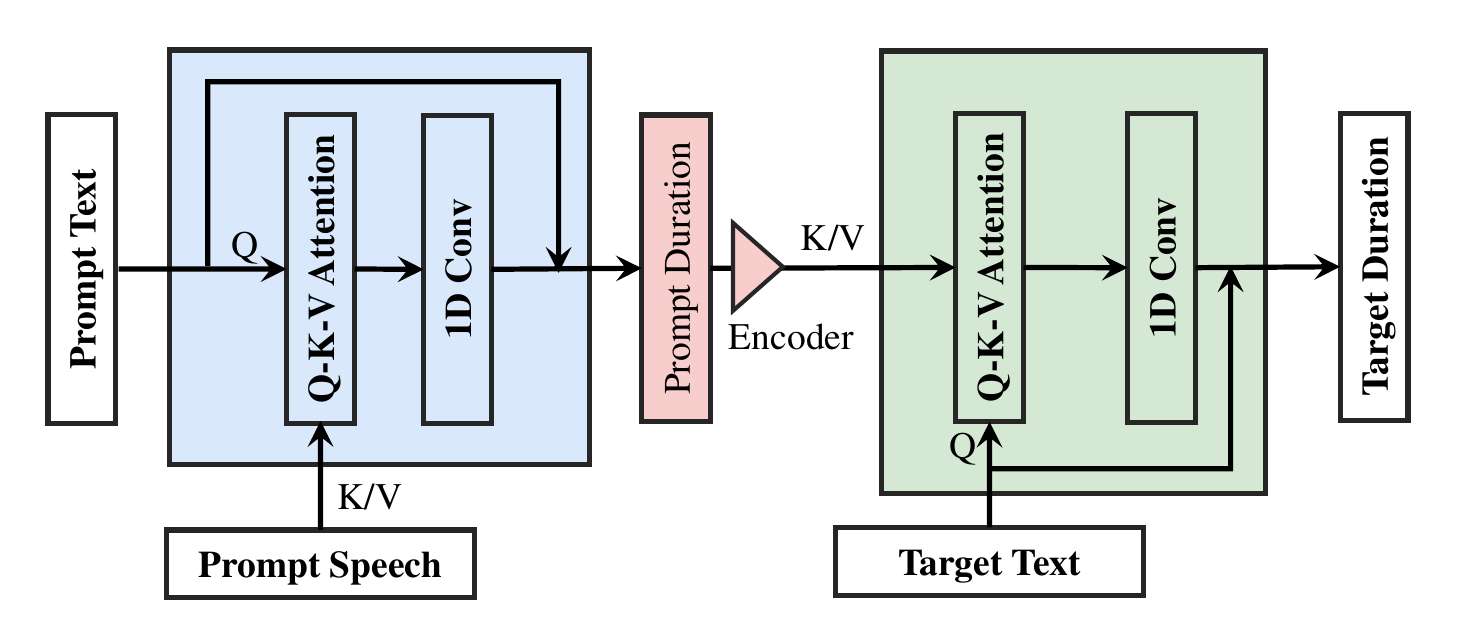}
    \caption{The process of obtaining target duration tokens from target text tokens and prompt speech tokens is depicted in the following manner: the blue module represents the Prompt Duration Extractor, while the green module represents the Duration Predictor.}
    \label{figurejsp2}
\end{figure}

MobileSpeech, based on FastSpeech2, adopts an Encoder-Decoder architecture. Different prompt strategies were designed in both stages to encourage mobileSpeech to capture the prosodic and timebre information of the prompt speech. During the experimental process, we observed that MobileSpeech generated speech with relatively uniform prosody and the distribution of durations exhibited excessive uniformity. Therefore, we abandoned the coarse-grained speech prompt guidance before the duration predictor. As illustrated in Figure \ref{figurejsp2}, the Duration Extractor incorporates a Q-K-V attention layer to integrate different information. It first extracts fine-grained prompt duration values from the prompt acoustic token and feeds them into the prompt duration encoder along with positional information. Subsequently, the prompt duration is used to guide the prediction of the target text's duration. Subsequent ablation experiments have confirmed that this approach can improve the occurrence of disfluent speech phenomena. Both the Duration Extractor and the Duration Predictor modules follow the same structure. The text is first passed through a normalization layer and then mapped to a Query. The Query, along with the prompt acoustic or prompt duration, undergoes a Q-V-K attention layer and a convolutional layer to obtain latent features. The latent features are then added to the text through residual connections to yield the corresponding output results.

In the Decoder stage of MobileSpeech, as described in Equation \ref{equation1} and \ref{equaton2}, during the training process, we randomly select a sequence of length k from the target text and speech to form a text-independent speech prompt. It is worth noting that, to ensure the robustness of the model, the speaker prompt in the Encoder and the speaker prompt in the Decoder are not identical sequences. In the Speech Mask Codec Decoder (SMD), all tokens of the input prompt acoustic in SMD are unmasked, and inspired by \cite{unicats} , we do not include corresponding positional information and prompt encoder in order to preserve the original features of the acoustic tokens as one of the inputs to the SMD module. This is explicitly highlighted with a blue line in Figure \ref{figurejsp1} (b). Additionally, in the SMD module, we pass the acoustic tokens through the prompt encoder and apply cross-attention to allow the target text, as a weak condition token, to incorporate some information from the prompt acoustic tokens, thereby maintaining contextual capabilities.

\subsection{Train and Inference}
During the training process, the Duration Predictor and Duration Extractor are optimized using the mean square error (MSE) loss, where the extracted duration serves as the training target. We denote the losses at Duration Predictor and Duration Extractor as $\mathcal L_{dur}$ and 
$\mathcal L_{promptdur}$, respectively. To improve alignment accuracy and reduce the information gap between the model input and output, we employ the Montreal forced alignment (MFA) \cite{mfa} tool to extract phoneme durations. The duration values are predicted in the logarithmic domain, which promotes a more Gaussian distribution and facilitates training. The SMD module is optimized using the cross-entropy loss function, considering both the first channel and the discrete acoustic tokens randomly selected under the Prank strategy. We use $\mathcal L_{smd}^{1}$ and 
$\mathcal L_{smd}^{j}$ to represent the losses of the first channel and the $j$-th channel, respectively. Where $2\le j\le N$. Thus, the overall loss $\mathcal L$ can be expressed as follows:
\begin{equation}
    \mathcal L=\mathcal L_{smd}^{1}+\mathcal L_{smd}^{j}+\mathcal L_{promptdur}+\mathcal L_{dur}
\end{equation}

During the inference stage, we iteratively generate discrete acoustic tokens by incorporating the prompt speech into the original text state. For the generation of the first channel of acoustic tokens, we employ the confidence-based sampling scheme proposed by \cite{maskgit,soundstorm} . Specifically, we perform multiple forward passes, and at each iteration $i$, we sample candidates for the masked positions. We then retain $P_{i}$ candidates based on their confidence scores, where $P_{i}$ follows a cosine schedule. As for the generation of the remaining channels, we directly select the token with the highest probability using a greedy strategy.

\section{Experiments}
\subsection{Experiment Setup}
\textbf{Datasets}
We trained MobileSpeech on datasets of varying scales and languages, and evaluated its performance on datasets comprising speech prompts of different lengths and environments. This allowed us to measure the effectiveness of MobileSpeech. For the English section, we trained MobileSpeech exclusively on the libriTTS \cite{libritts} , which comprises 580 hours of data from 2,456 speakers. For the Mandarin Chinese section, MobileSpeech was trained on an internal dataset totaling 40,000 hours. 
\\
\textbf{Baselines}
We consider the latest SOTA models as our baselines. For the English segment, we utilize YourTTS \cite{yourtts} \footnote{\url{https://github.com/Edresson/YourTTS}}in its official version. We have successfully reproduced results close to the official ones for VALL-E \cite{valle}, surpassing the reproduced results found online \footnote{\url{https://github.com/enhuiz/vall-e}}. As for NaturalSpeech2 \cite{naturalspeech2}, we employ the reproduced version available online. Additionally, we have achieved results similar to the official ones for Megatts \cite{megatts}. For the Chinese segment, we compare our results with the open-source interface \footnote{\url{https://www.volcengine.com/product/voicecloning}} of the current state-of-the-art Chinese model Megatts2 (the version of the interface is November 2023) \cite{megatts2}.
\\
\textbf{Training and Inference Settings}
MobileSpeech was trained on 40,000 hours of Mandarin Chinese data using 32 NVIDIA A100 40G GPUs. The discrete codec was trained end-to-end on the training dataset, with the encoder part of Encodec \cite{encodec} and the decoder part of Vocos \cite{vocos} being optimized through engineering modifications. For the 580-hour English training data, the discrete codec employed a pre-trained version. MobileSpeech was trained for 12 epochs on 4 NVIDIA A100 40G GPUs, with each batch accommodating 3500 frames of the discrete codec. We optimized the models using the AdamW optimizer with parameters $\beta _{1}$ = 0.9 and $\beta_{2}$ = 0.95. The learning rate was warmed up for the first 5k updates, reaching a peak of $5\times 10^{-4}$, and then linearly decayed.
\\
\textbf{Automatic metrics}
We have aligned all our objective experiments with VALL-E \cite{valle}. To evaluate speaker similarity (SPK) between the original prompt and synthesized speech, we employ WavLM-TDNN \cite{wavlm}. However, due to updates in the repository, we have updated the feature extractor, but all our models have been tested by using the same metrics. For assessing automatic speech recognition (ASR) performance, we conduct ASR on the generated audio and calculate the word error rate (WER) compared to the original transcriptions. In this experiment, we utilize the HuBERT-Large \cite{hubert} model fine-tuned on LibriSpeech 960h as the ASR model. This model is a CTC-based model without language model fusion. To compute the real-time factor, we measure the ratio of the total length of generated speech to the total time taken by the model to generate the corresponding speech features . The inference time in the vocoder component will not be taken into account. The calculation is performed on an empty A100 GPU. 
\begin{table*}[t]
\centering
\begin{adjustbox}{width=\textwidth}
\begin{tabular}{c|ccccccc}
\hline
Model & Data & WER $\downarrow$  & SPK $\uparrow$ & RTF $\downarrow$  & MOS-Q $\uparrow$ & MOS-P $\uparrow$ & MOS-S $\uparrow$ \\
\hline
GT Codec & - &2.4 & 0.871 &- &4.41$\pm$0.08 & 4.28$\pm$0.10 & 4.45$\pm$0.06\\ 
\hline
YourTTS & 640 & 7.7 & 0.504 & 0.22 &3.71$\pm$0.11 & 3.59$\pm$0.14 & 3.68$\pm$0.12\\ 
VALL-E & 60000 & 5.9 & \textbf{0.751} & 0.94 & \textbf{4.22$\pm$0.06} & 4.09$\pm$0.13 & \textbf{4.16$\pm$0.07}\\ 
VALL-E-Continue& 60000 & 3.8 & 0.734 &0.94 &4.12$\pm$0.12 & \textbf{4.13$\pm$0.10} & 4.11$\pm$0.11\\ 
NaturalSpeech2-Continue& 580 &4.6&0.581& 0.35 &3.85$\pm$0.11 & 3.69$\pm$0.14 & 3.76$\pm$0.08\\ 
MegaTTS-Continue& 580 & 5.8 &0.615 &0.39 &3.93$\pm$0.08 & 3.85$\pm$0.09 & 3.89$\pm$0.09\\ 
MobileSpeech-Continue& \textbf{580} & \textbf{3.1} & \textbf{0.688} & \textbf{0.09} &\textbf{4.06$\pm$0.07} & \textbf{4.02$\pm$0.08} & \textbf{4.05$\pm$0.10}\\ 
\hline
\end{tabular}
\end{adjustbox}
\caption{The results of different zero-shot TTS models on the LibriSpeech test-clean dataset.}
\label{table1}
\end{table*}
\textbf{Human evaluation}
We conduct the MOS (mean opinion score) and CMOS (comparative mean opinion score) evaluation on the test set to measure the audio naturalness via crowdsourcing. We keep the text content and prompt speech consistent among different models to exclude other interference factors. We randomly choose 50 samples from the test set of each dataset for thesubjective evaluation and each audio is listened to by at least 20 testers. We analyze the MOS in three aspects: MOS-Q (Quality: clarity, high-frequency, and original timbre reconstruction), MOS-P (Prosody: naturalness of pitch, energy, and duration), and MOS-S (Speaker similarity). We also analyze the CMOS in terms of audio quality and speech prosody. See Appendix \ref{appendixd} for details.
\subsection{LibriSpeech Evaluation}

\begin{table}
\centering
\begin{adjustbox}{width=0.5\textwidth}
\begin{tabular}{c|cccc}
\hline
\textbf{Model} & CMOS-Q $\uparrow$ & CMOS-P $\uparrow$ & CMOS-S $\uparrow$ \\
\hline
GT codec & +0.07 & +0.15 & +0.03\\
MegaTTS 2& -0.16 & -0.05 & -0.24\\ 
MobileSpeech& \textbf{0.00} & \textbf{0.00} &\textbf{0.00}\\ 
\hline
\end{tabular}
\end{adjustbox}
\caption{The CMOS (comparative Mean Opinion Score) results of different models on the Chinese Mandarin test set.}
\label{table2}
\end{table}

Regarding the inference phase, to ensure fair comparisons, we followed the experimental protocols outlined in VALL-E \cite{valle} and employed the LibriSpeech test-clean dataset \cite{librispeech} , ensuring no overlap with our training data. We specifically utilized samples from the LibriSpeech test-clean dataset with durations ranging from 4 to 10 seconds, resulting in a subset of 2.2 hours. Following VALL-E approach, we use the whole transcription and the first 3 seconds of the utterance as the phoneme and acoustic prompts respectively, and ask the model to generate the continuations, the continue setting of Table \ref{table1}  is referred to this.

By observing Table \ref{table1}, the following conclusions can be drawn: (1) Under the same scale of 580 hours training data of libriTTS (where yourtts is trained on a combined dataset of VCTK , LibriTTS , and TTS-Portuguese \cite{ttsdata}), MobileSpeech significantly outperforms YourTTS in terms of SPK and MOS, surpassing the current SOTA models NaturalSpeech2 and MegaTTS. This indicates that MobileSpeech is capable of generating higher audio quality, better prosody, and more similar audio. The outstanding results are attributed to the utilization of the baseline generated by the MASK parallel codec and the additional design of the SMD module, Speaker Prompt module in MobileSpeech. (2) In terms of audio quality, MobileSpeech also achieves comparable results to VALL-E, which was trained on a large-scale dataset of 60,000 hours, with a speaker similarity metric difference of only 0.046. This difference is attributed to the disparity in the training data volume. (3) MobileSpeech achieves the best results in terms of robustness measured by the Wer metric, even surpassing VALL-E by 0.7. This is because the autoregressive structure of VALL-E often suffers from cumulative errors, resulting in lower robustness. Our experiments reveal that increasing the amount of data can alleviate the issue of high robustness. (4) MobileSpeech exhibits remarkable performance in terms of the Real-Time Factor (RTF) metric, as it can generate 10 seconds of audio within one second. Compared to VALL-E, MobileSpeech achieves an 11-fold speed improvement, and compared to MegaTTS and NaturalSpeech2, it achieves a 4-fold speed improvement. Although YourTTS achieves an RTF of 0.22, its simple structural design leads to unsatisfactory performance in terms of audio quality, speaker similarity, and model robustness.

\subsection{Mandarin Evaluation}
As our inference dataset for Mandarin Chinese, we compiled a collection of 200 real-world mobile-end recordings with arbitrary durations ranging from 3 to 10 seconds, representing a mixture of clean and noisy scenarios. This dataset serves as our Mandarin Chinese test set. We employed CMOS to measure the performance of MobileSpeech and the current SOTA model MegaTTS2 on the Mandarin Chinese dataset. Considering that MegaTTS2 is a paid model, we incurred significant costs to conduct this experiment.

The experimental results, as shown in Table \ref{table2}, demonstrate that MobileSpeech outperforms MegaTTS2 comprehensively in terms of MOS-Q, MOS-P, and MOS-S under various conditions, including large-scale training corpora, different language categories, varying prompt lengths, and noisy speech prompts. Particularly, MobileSpeech achieves a significantly higher score of 0.24 in terms of speaker similarity compared to MegaTTS2. By conducting a cross-comparison between MobileSpeech and GT Codec in terms of CMOS-Q, CMOS-P, and CMOS-S metrics, we can observe that MobileSpeech exhibits superior performance in terms of speaker similarity but slightly lags behind in audio prosody.

\begin{table}
\centering
\begin{tabular}{c|cc}
\hline
\textbf{Model} & WER $\downarrow$ & SPK $\uparrow$   \\
\hline
MobileSpeech w/o durprompt& 3.6 & 0.638 \\ 
MobileSpeech w/o onechannel& 4.2 & 0.614\\
MobileSpeech& \textbf{3.1} & \textbf{0.688}\\ 
\hline
\end{tabular}
\caption{The ablation experiments of the SMD module and the Speaker Prompt module were conducted to test SPK and WER.}
\label{table3}
\end{table}

\begin{table}
\centering
\begin{adjustbox}{width=0.5\textwidth}
\begin{tabular}{c|ccc}
\hline
\textbf{Model} & MOS-Q $\uparrow$ & MOS-P $\uparrow$ & MOS-S $\uparrow$ \\
\hline
MobileSpeech w/o durprompt &4.03$\pm$0.05 & 3.90$\pm$0.11 & 3.97$\pm$0.12 \\ 
MobileSpeech w/o onechannel &3.98$\pm$0.12 & 3.95$\pm$0.09 & 3.92$\pm$0.11 \\
MobileSpeech&\textbf{4.06$\pm$0.07} & \textbf{4.02$\pm$0.08} & \textbf{4.05$\pm$0.10} \\ 
\hline
\end{tabular}
\end{adjustbox}
\caption{The ablation experiments of the SMD module and the Speaker Prompt module were conducted to test MOS.}
\label{table4}
\end{table}

\subsection{Parameters Evaluation}
To begin, we present a comprehensive account of the parameters for all state-of-the-art models. In the VALL-E framework, the AR model comprises 229 million parameters, while the NAR model consists of 235 million parameters, resulting in a total of 465 million parameters. In the NaturalSpeech2 model, the overall parameter count is 435 million. Within the MegaTTS system, the VQ-GAN model encompasses 243 million parameters, the LLM model comprises 124 million parameters, and the hifigan vocoder consists of 106 million parameters, resulting in a cumulative parameter count of 473 million. As outlined in the accompanying figure, the Mobilespeech model has a total parameter count of 207 million (detailed parameters can be found in Appendix \ref{appendixb} of the paper), demonstrating its status as the most lightweight option.

\subsection{Ablation experiment}

We conducted ablation experiments on the LibriSpeech test-clean dataset. Firstly, we omitted fine-grained prompt duration and directly used prompt acoustic tokens to guide the generation of target text. Additionally, in the SMD module, prompt acoustic tokens were solely used as the input of SMD. This version was referred to as MobileSpeech w/o durprompt. For the MobileSpeech w/o onechannel version, we replaced the SMD module with randomly selecting a layer for mask prediction. The results are presented in Tables \ref{table3} and \ref{table4}. It can be observed that both MobileSpeech w/o durprompt and MobileSpeech w/o onechannel exhibited certain decreases in WER, SPK, and MOS values compared to MobileSpeech. This suggests that each component of MobileSpeech's design is necessary. During our experimental process, we found that the MobileSpeech w/o durprompt version exhibited a certain amount of disfluency, which affected the overall prosody of the audio. This indicates that fine-grained duration prompts indeed contribute to better modeling of duration information.

We also conducted ablation experiments on the number of iterations in the first channel of MobileSpeech, which has a significant impact on the inference speed. As shown in Table \ref{table5}, we observed that the number of iterations is inversely proportional to the real-time factor (RTF), and blindly increasing the number of iterations does not necessarily improve the model's performance. However, when the number of iterations is reduced to 1, the model exhibits significant degradation in its capabilities. Consequently, we selected 8 iterations as the final choice, striking a balance between performance and efficiency.

\begin{table}
\centering
\begin{tabular}{cccc}
\hline
\textbf{iterations}  &  RTF $\downarrow$ &WER $\downarrow$ &SPK $\uparrow$  \\
\hline
24 & 0.20 & 3.1 &0.698\\ 
16 & 0.15 & 3.0 &0.696\\ 
\textbf{8}& \textbf{0.09} & \textbf{3.1}& \textbf{0.688} \\ 
4& 0.07 & 4.2 &0.615\\ 
1& 0.05 & 5.7 &0.483\\ 
\hline
\end{tabular}
\caption{The impact of different iterations on the first channel of MobileSpeech.}
\label{table5}
\end{table}

\section{Conclusion}
In this paper, we propose MobileSpeech, a Fast and Lightweight Framework for Mobile zero-shot Text-to-Speech. MobileSpeech achieves a unique combination of audio naturalness and similarity performance, comparable to other zero-shot TTS systems on English corpora, while significantly enhancing inference speed through its distinctive SMD and Speaker modules. MobileSpeech stands as the first real-time zero-shot TTS system deployable at the edge, demonstrating state-of-the-art performance on Chinese corpora through extensive training on large-scale Chinese language data. We not only validate the significance of each module in MobileSpeech through ablation experiments but also analyze the effects of different configurations and iterations to adapt to various deployment scenarios in mobile environments. 

 \section{Acknowledgments}
 This work was supported in part by the National Key R\&D Program of China under Grant No.2022ZD0162000

\section{Limitations and Future Work}
In this section, we will analyze the limitations of MobileSpeech and discuss potential future work.

\textbf{Failure cases} During our experiments in real-world scenarios (Figure \ref{table2}), we observed occasional instances of stuttering and unstable pitch in MobileSpeech. This could be attributed to noisy input prompts or the possibility that the mask mechanism excessively focuses on local positions. In future work, we aim to enhance the model's noise robustness and design a more robust mask mechanism that considers a broader context.

\textbf{New Tasks:} The current zero-hhot TTS task itself has significant potential for expansion. Most existing models primarily focus on similarity in timbre. However, future research can extend zero-shot TTS to include similarity in rhythm, emotion, and language. It could even involve using multiple audio segments to control different parts or combining with text style control, representing meaningful new tasks.

\textbf{Broader impacts:} Since MobileSpeech could synthesize speech that maintains speaker identity, it may carry potential risks in misuse of the model, such as spoofing voice identification or impersonating a specific speaker. To mitigate such risks, it is possible to build a detection model to discriminate whether an audio clip was synthesized by MobileSpeech.

\bibliography{custom}

\appendix

\section{Discrete Acoustic Codec}
\label{appendixA}
Since audio is typically stored as a sequence of 16-bit integer values, a generative model is required to output $2^{16}=65536$ probabilities per timestep to synthesize the raw audio. In addition, the audio sample rate exceeding ten thousand leads to an extraordinarily long sequence length, making it more intractable for raw audio synthesis. In these days, neural acoustic codec \cite{soundstream,encodec,hificodec,vocos}, with their ability of reconstructing high-quality audio at very low bitrates, sub-sequently allowed for extending discrete modeling to audio signals as diverse as multi-speaker speech. To be specific, Soundsteam \cite{soundstream} which relies on a model architecture composed by a fully convolutional encoder/decoder network and a residual vector quantizer (RVQ) can efficiently compress speech. Encodec \cite{encodec} consists in a streaming encoder-decoder architecture with quantized latent space trained in an end-to-end fashion.Hificodec \cite{hificodec} propose a group-residual vector quantization (GRVQ) technique To reduce the number of quantizers. Vocos \cite{vocos} closing the gap between time-domain and fourier-based neural vocoders for high-quality audio synthesis. After our detailed experiments, MobileSpeech finally selected encodec and vocos as the encoder and decoder of acoustic token.

\section{Codec Modeling}
\label{appendixB}
\begin{figure}[htbp]
\centering
\includegraphics[height=9cm, width=8cm]{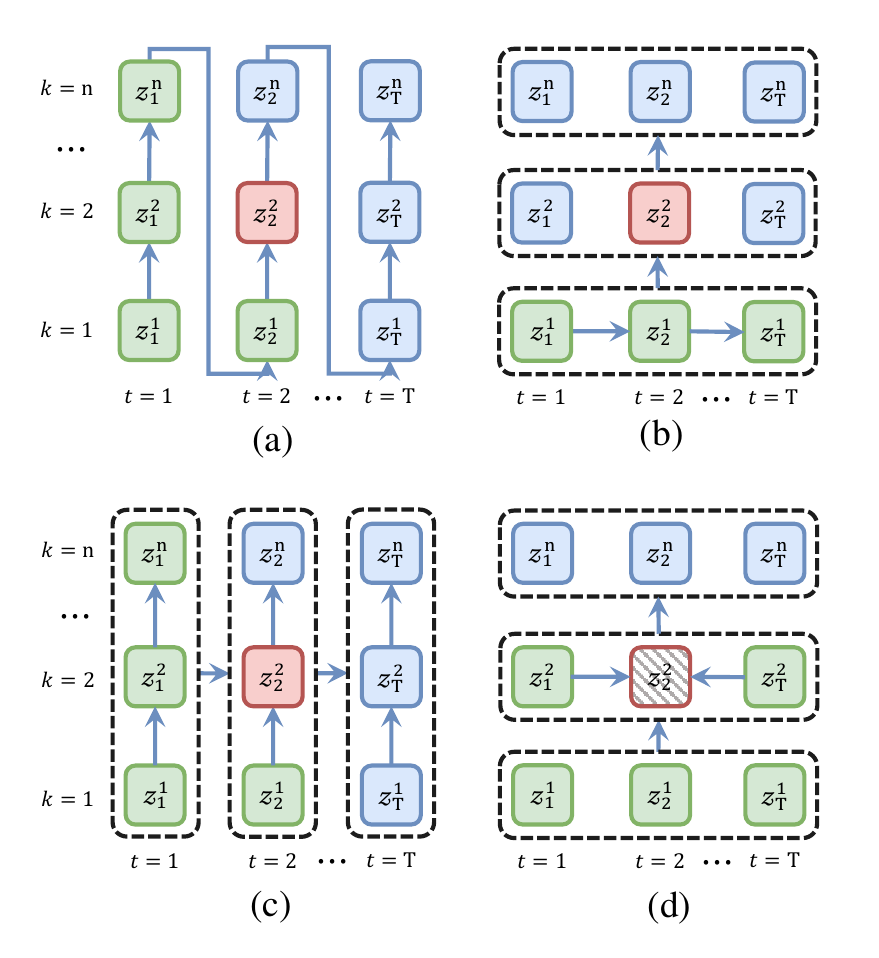}
\caption{Based on the RVQ structure, different articles adopt distinct modeling approaches for discrete codecs, where (a) SpearTTS \cite{speartts}, (b) VALL-E \cite{valle}, (c) Uniaudio \cite{uniaudio} and (d) MobileSpeech.}
\label{figurejsp3}
\end{figure}

In this section, we provide a detailed comparison of different modeling approaches for the RVQ structure based on discrete codecs. As depicted in Figure \ref{figurejsp3}, (a) represents a fully sequential unfolding of each frame's channels followed by autoregressive modeling, which incurs significant time overhead as the number of channels increases. (b) involves splitting the codec generation into two stages, where the first stage autoregressively generates the codec for the first channel, and the second stage concurrently generates the codecs for the remaining channels. (c) adopts global attention in the temporal dimension and local attention in the channel dimension for modeling. All three methods (a), (b), and (c) are fundamentally autoregressive models, but due to limitations in inference speed and memory consumption, they are not suitable for deployment on mobile devices. In contrast, MobileSpeech adopts a mask-based parallel generation approach, offering a potential deployment solution.

\section{Model Structure}
\label{appendixb}
In this section, we introduce the detailed structure of proposed MobileSpeech. To be specific, we will introduce the architecture of text encoder, prompt duration encoder, prompt encoder, cross-attention and ConFormer \cite{conformer} used in MobileSpeech. Text encoder and prompt duration encoder share the same architecture which is comprised of several FFT blocks as used by FastSpeech2 \cite{fastspeech2} . The value of probability parameter $\alpha  $ in MobileSpeech is $0.6$ . The detailed configuration information is shown in the table \ref{table6}.

\begin{table*}[t]
\centering
\begin{tabular}{{c}|{l}{c}|{c}}
    \hline
    \multicolumn{1}{c|}{\textbf{Module}} & \multicolumn{2}{c|}{
        \textbf{Hyper-parameters}
    } & \multicolumn{1}{c}{\textbf{Parameters Number}} \\
    \hline
    \multirow{4}{*}{Prompt Text Encoder} 
    & Number of FFT Blocks & 2 & \multirow{4}{*}{20.994M} \\
    & Conv1D kernel size & 3 & \\ & Conv1D input channels & 512 & \\
    & Attention heads & 8 & \\
    \hline
    \multirow{4}{*}{Text Encoder} & Number of FFT Blocks & 4 & \multirow{4}{*}{41.988M} \\
    & Conv1D kernel size & 3 & \\ & Conv1D input channels & 512 & \\
    & Attention heads & 8 & \\
    \hline
    \multirow{5}{*}{Prompt Encoder} & Number of layers & 2 & \multirow{5}{*}{16.243M} \\
    & Hidden channels & 512 & \\
    & Conv1D kernel size& 3 & \\ & Filter channels & 2048 & \\
    & Attention heads & 4 & \\
    \hline
    \multirow{3}{*}{Duration Extractor} & Number of attention layers & 10 & \multirow{3}{*}{19.173M} \\
    & Hidden channels & 512 & \\ 
    & Attention heads & 8 & \\
    \hline
    \multirow{3}{*}{Duration Predictor} & Number of attention layers & 10 & \multirow{3}{*}{19.173M} \\
    & Hidden channels & 512 & \\ 
    & Attention heads & 8 & \\
    \hline
    \multirow{4}{*}{Prompt Duration Encoder} & Number of FFT Blocks & 2 & \multirow{4}{*}{20.994M} \\
    & Conv1D kernel size & 3 \\ & Conv1D input channels & 512 & \\
    & Attention heads & 8 & \\
    \hline
    \multirow{5}{*}{Conformer} & Number of Encoder layers & 16 & \multirow{5}{*}{67.200M} \\
    & Attention dim & 512 & \\
    & Attention heads & 16 & \\
    & Linear units & 1024 & \\
    & Positionwise Conv1D Kernel & 5 & \\
    & Dropout probability& 0.1 \\
    \hline
\end{tabular}
\caption{The configuration of modules of MobileSpeech.}
\label{table6}
\end{table*}

\section{Algorithm}
\label{appendixc}

We provide a detailed description of the $P_{rank}$ algorithm in the Algorithm 1 provided below.

\begin{table}[htbp]
\begin{adjustbox}{width=\textwidth}
\begin{tabular}{l} 
    \hline
    \textbf{Algorithm 1 } $P_{rank}$ algorithm \\
    \hline
    \quad \textbf{Require:} The channels set ${\{c_i\}_{i=2}^{N}}$ and decreasing corresponding weights ${\{w_{i}}\}_{i=2}^{N}$;\\
    \quad where $w_{j}> w_{j+1}, j\in \left \{ 2,3,\cdots ,N-1 \right \} $ \\
    \quad \textbf{Initialize} $\mathcal{C}_U=\emptyset$; \\
    \quad \textbf{for} each $c_i$ \textbf{do}\\
    \qquad Repeat $c_i$ for $w_{i}$ times; \\
    \qquad Add repeated $c_i$ to $\mathcal{C}_U$; \\
    \quad \textbf{end for} \\
    \quad Random sample a channel $c$ from $\mathcal{C}_U$;\\
    \hline
\end{tabular}
\end{adjustbox}
\end{table}

\section{MOS Evaluation}

Below, we provide a detailed explanation of the requirements and standards we employed for measuring Mean Opinion Score (MOS). 














\begin{figure}[htbp]
    \centering
    \includegraphics[height=10cm, width=8cm]{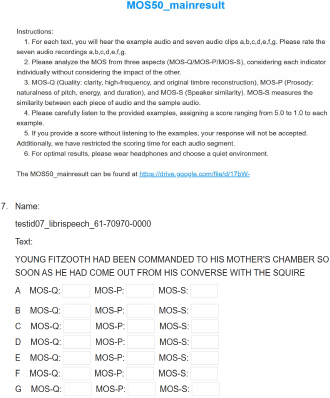}
    \caption{MOS evaluation procedure}
    \label{figurejsp5}
\end{figure}

\label{appendixd}

\end{document}